\documentclass[runningheads]{llncs}
\usepackage{graphicx}
\usepackage{courier}
\usepackage{multirow}
\usepackage{rotating}
\usepackage{etoolbox}
\usepackage{url}
\usepackage{amsmath}
\usepackage{kantlipsum} % just for the example
\makeatletter
\renewcommand*{\@fnsymbol}[1]{\ensuremath{\ifcase#1\or *\or \dagger\or \ddagger\or
    \mathsection\or \mathparagraph\or \|\or **\or \dagger\dagger
    \or \ddagger\ddagger \else\@ctrerr\fi}}
\makeatother
\begin{document}
\title{Brain tumour segmentation using a triplanar ensemble of U-Nets on MR images}
\titlerunning{Brain tumour segmentation using a triplanar ensemble of U-Nets}
\author{Vaanathi Sundaresan\inst{1}\orcidID{0000-0002-9451-4779}\thanks{Corresponding~author; \url{https://www.ndcn.ox.ac.uk/team/vaanathi-sundaresan}, Email~id:~vaanathi.sundaresan@ndcn.ox.ac.uk} \and
Ludovica Griffanti\inst{1,2}\orcidID{0000-0002-0540-9353}\thanks{Contributed equally to this work} \and
Mark Jenkinson\inst{1,3,4}\orcidID{0000-0001-6043-0166}$^\dagger$}

% \author{Vaanathi Sundaresan\inst{1} \and
% Mark Jenkinson\inst{1} \and
% Giovanna Zamboni\inst{1,2} \and
% Peter Rothwell\inst{1,2} \and
% Robert Dineen\inst{3} \and
% Dorothee Auer\inst{3} \and
% Ludovica Griffanti\inst{1}}
% Vaanathi Sundaresan ( University of Oxford) <vaanathi.sundaresan@some.ox.ac.uk>
% Mark Jenkinson ( University of Oxford ) <mark.jenkinson@ndcn.ox.ac.uk>
% Giovanna Zamboni ( University of Oxford) <giovanna.zamboni@ndcn.ox.ac.uk>
% Peter Rothwell ( University of Oxford) <peter.rothwell@ndcn.ox.ac.uk>
% Robert Dineen ( University of Nottingham) <rob.dineen@nottingham.ac.uk>
% Dorothee Auer ( University of Nottingham) <dorothee.auer@nottingham.ac.uk>
% Ludovica Griffanti ( University of Oxford) <ludovica.griffanti@ndcn.ox.ac.uk>
%
\authorrunning{V. Sundaresan et al.}

\institute{Wellcome Centre for Integrative Neuroimaging, Oxford Centre for Functional MRI of the Brain,
Nuffield Department of Clinical Neurosciences, University of Oxford, UK \and
Wellcome Centre for Integrative Neuroimaging, Oxford Centre for Human Brain Activity,
Department of Psychiatry, University of Oxford, Oxford, UK \and
Australian Institute for Machine Learning (AIML), School of Computer Science, The University of
Adelaide, Adelaide, Australia \and 
South Australian Health and Medical Research Institute (SAHMRI), Adelaide, Australia}
% \institute{FMRIB, Wellcome Centre for Integrative NeuroImaging, Nuffield Department of Clinical Neurosciences, University of Oxford, UK \and
% Centre for Prevention of Stroke and Dementia, Nuffield Department of Clinical Neurosciences, University of Oxford, UK  \and
% NIHR Nottingham Biomedical Research Centre; University of Nottingham, Nottingham, UK }

\maketitle   

\begin{abstract}
Gliomas appear with wide variation in their characteristics both in terms of their appearance and location on brain MR images, %. Accurate automated detection of gliomas would be useful for timely diagnosis and treatment planning. However, the intrinsic histological variations and diversity in their biological origin 
which makes robust tumour segmentation highly challenging, and leads to high inter-rater variability even in manual segmentations. In this work, we propose a triplanar ensemble network, with an independent tumour core prediction module, for accurate segmentation of these tumours and their sub-regions. On evaluating our method on the MICCAI Brain Tumor Segmentation (BraTS) challenge validation dataset, for tumour sub-regions, we achieved a Dice similarity coefficient of 0.77 for both enhancing tumour (ET) and tumour core (TC). In the case of the whole tumour (WT) region, we achieved a Dice value of 0.89, which is on par with the top-ranking methods from BraTS'17-19. Our method  %achieved the 5$^{\text{th}}$ highest evaluation score and 
achieved an evaluation score that was the equal 5$^{th}$ highest value (with our method ranking in 10$^{th}$ place) in the BraTS'20 challenge, with mean Dice values of 0.81, 0.89 and 0.84 on ET, WT and TC regions respectively on the BraTS'20 unseen test dataset. 

\end{abstract}

\setlength{\parindent}{0.5cm}
\section{Introduction}
\label{sec:intro}

Gliomas, the most common class of brain tumours, occur with different levels of aggressiveness with highly heterogeneous sub-regions including  invaded edematous tissue or peritumoral edema (ED)  and tumour core region including necrotic core (NCR), non-enhancing tumour (NET), and enhancing tumour (ET) \cite{menze2015multimodal}, \cite{bakas2018indentifying}. Accurate and reproducible automated detection of gliomas would aid in the timely diagnosis and staging of tumours in a clinical setting, and reliable analysis in large population studies. However, the intrinsic histological variations of gliomas are further complicated by the heterogeneity in characteristics (e.g. intensity) of tumours on MRI scans. Various sub-regions of gliomas occur with wide variations in their appearance and shape depending on their biological conditions, making their segmentation highly challenging and often leading to high inter-rater variability even in expert clinicians’ segmentations across different datasets \cite{menze2015multimodal}.

\medskip
The MICCAI Brain Tumor Segmentation (BraTS) challenges aim to provide accurate segmentation of brain tumours on multimodal MR images \cite{menze2015multimodal}, \cite{bakas2018indentifying}, \cite{bakas2017advancing}, \cite{bakas2017segmentation1}, \cite{bakas2017segmentation}. Several methods, including recent deep learning methods, have been proposed in BraTS challenges. The top ranking methods used convolutional neural network (CNN) architectures \cite{pareira2016brain}, \cite{kamnitsas2017efficient}, \cite{havaei2017brain}, \cite{shen2017boundary} mostly using ensemble networks \cite{kamnitsas2017ensembles},  \cite{mckinley2018ensembles}, \cite{mckinley2019triplanar} and/or encoder-decoder frameworks \cite{yang2017automatic}, \cite{myronenko20183d}. U-Nets \cite{ronneberger2015u}, one of the most popular encoder-decoder networks, were used successfully with accurate results for tumour segmentation \cite{kim2017brain}, \cite{isensee2017brain}, \cite{isensee2018no}, \cite{jiang2019two}. Regarding the model dimensions, both 2D \cite{pareira2016brain} and 3D networks \cite{kamnitsas2017efficient} were used with additional post-processing steps (e.g. conditional random fields used in \cite{kamnitsas2017efficient}, \cite{shen2017boundary}) and occasionally within multi-scale architectures \cite{kamnitsas2017efficient}, \cite{havaei2017brain} and multi-step cascaded frameworks \cite{mckinley2019triplanar}, \cite{jiang2019two}. Some methods aimed to leverage the advantages of both 2D and 3D architecture by using triplanar ensembles of CNNs \cite{mckinley2019triplanar}, providing accurate segmentation with fewer parameters than 3D networks. Further, modifications to the loss functions have been proposed \cite{shen2017boundary}, \cite{mckinley2019triplanar} for overcoming class imbalance, reliable tumour core detection and accurate segmentation of tumour boundaries.

\medskip
We propose a fully automated deep learning method for brain tumour segmentation using a triplanar ensemble architecture consisting of a 2D U-Net in each plane (axial, sagittal and coronal) of MR images. Our method uses a combination of loss functions in order to overcome the class imbalance and includes an independent tumour core prediction module to refine the segmentation of tumour core sub-regions. We study the effect of various components of our architecture on the segmentation results by performing an ablation study. We evaluate our method on BraTS’20 training and validation datasets, which exhibit wide heterogeneity in tumour characteristics and provides a benchmark to assess the robustness of our segmentation method. Finally, the results of our method on BraTS'20 test dataset shows that our method provides accurate segmentation of tumour regions, ranking among the top 10 best performing methods of the challenge.

\section{Materials and methods}
\label{sec:method}

\subsection{Data}
\label{ssec:data}
We evaluated the performance of the proposed method on the publicly available BraTS'20 dataset, consisting of pre-operative multimodal MRI scans, with 369 training cases. For each subject, the given input modalities include FLAIR, T1-weighted (T1), post-contrast T1-weighted (T1-CE) and T2-weighted (T2) images.  The manual segmentations for the training dataset consists of 3 labels \cite{menze2015multimodal}, \cite{bakas2018indentifying}: NCR/NET, ED and ET. The input modalities (FLAIR, T1, T1-CE, T2) were already co-registered to the same anatomical template of dimension 240 $\times$ 240 $\times$ 155, interpolated to the same resolution (1 mm$^3$ isotropic) and skull-stripped. In addition to the training data, 125 validation cases were %provided (without manual segmentations) for the initial testing of our method.
provided without manual segmentations (referred from now on as ``unlabelled validation dataset") to enable an initial validation of the method via the challenge's online evaluation platform (using the ground truth at their end). Additionally, 166 test cases without manual segmentations (referred from now on as ``unseen test dataset") were released for a duration of 48 hours for the final testing via the online evaluation platform (again, using their local ground truth).

\subsection{Preprocessing}
\label{ssec:preprocess}

We cropped the images to a standard size of 192 $\times$ 192 $\times$ 160 voxels %(padding with zeros on both ends in the \textit{z}-direction), 
so that field of view (FOV) is close to the brain, and applied Gaussian normalisation to the intensity values. We then extracted 2D slices from the volumes from axial, sagittal and coronal planes with dimensions of 192 $\times$ 192, 192 $\times$ 160 and 192 $\times$ 160 voxels respectively.

\begin{figure*}[h!]
    \centering
    \includegraphics[width=\textwidth]{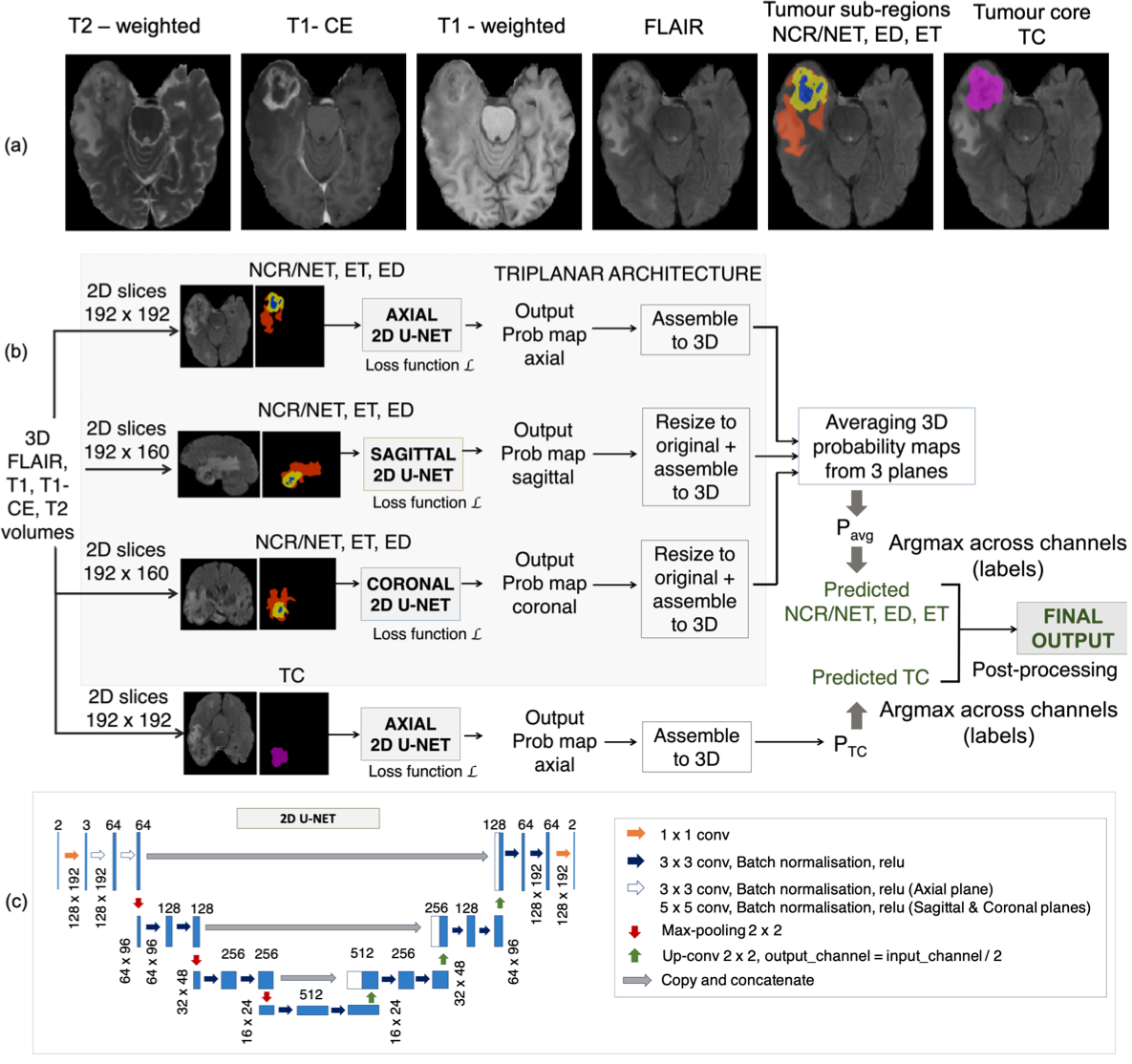}
    \caption[]{\footnotesize Proposed triplanar ensemble network architecture. (a) Input modalities in the axial plane along with manual segmentations for NCR/NET (blue), ET (yellow), ED (red) and TC (magenta), (b) the proposed network and (c) 3-layer deep U-Net blocks used in (b). Slices with 4 channels (input modalities) were provided to all U-nets.}
    \label{fig:network_arch}
\end{figure*}

\subsection{CNN architecture} 
\label{ssec:arch}

We used the triplanar architecture proposed in \cite{sundaresan2020triplanar}\footnote{The original tool using triplanar model is available at \url{https://git.fmrib.ox.ac.uk/vaanathi/truenet}.\\ The tool for tumour segmentation is available at \url{https://git.fmrib.ox.ac.uk/vaanathi/truenet_tumseg}}. Briefly, as shown in fig.~\ref{fig:network_arch}, the triplanar architecture consists of three 2D U-Nets, one for each plane, taking FLAIR, T1, T1-CE and T2 slices as input channels. On the training dataset, we observed that while the manual segmentation for the cumulative tumour core (TC, which is ET + NCR/NET) (fig.~\ref{fig:network_arch}a) was quite consistent, there were wide variations in those of individual sub-regions (NCR/NET and ET), due to their underlying histological heterogeneity. Therefore, in order to reduce the inconsistencies in the boundaries of ET and NCR/NET labels, %we considered TC sub-region directly in addition to the other 3 labels using the triplanar network, as shown in figure~\ref{fig:network_arch}. Hence for training and testing, 
we modified the ground truth labels to include TC in addition to the provided labels, obtaining 4 labels: ET, ED, NCR/NET and TC. During training, we used the ET, NCR/NET and ED labels to train the U-Nets in the triplanar architecture, while we used the TC label to train an independent axial U-Net as shown in fig.~\ref{fig:network_arch}b. We later used the TC regions in the post-processing step (refer section~\ref{ssec:postproc}) to further refine the final output labels (ET, NCR/NET and ED).

\medskip
We trimmed the depth of the classic U-Net \cite{ronneberger2015u} in each plane to a depth of 3-layers (fig.~\ref{fig:network_arch}b), to reduce the computational load. While axial U-Nets use 3 $\times$ 3 convolutional kernels in the initial layer, the other two U-Nets use 5 $\times$ 5 kernels. This helped to learn more generic lesion patterns, thus avoiding any discontinuities in segmentation along the z-dimension. In the ensemble model, we trained the U-Nets in each plane independently using 2D slices extracted in each plane. We used a combination of cross-entropy (CE) and Dice loss (DL) functions in order to overcome the effect of class imbalance between tumour/edema and healthy tissue. The loss function was computed batch-wise as shown below in eqn~\ref{eqn:loss_function}.

\begin{equation}
    L = CE + DL =  -\sum\limits_{c = 1}^{C} y_c(x)\text{log}(p_c(x)) - \frac{1}{C}\sum\limits_{c = 1}^{C}\frac{2 \times \sum_{x = 1}^N M_c(x)\cdot PL_{c}(x)}{\sum_{x= 1}^N M_c(x) + \sum_{x= 1}^N PL_{c}(x)} 
    \label{eqn:loss_function}
\end{equation}

where $p_c(x)$ denotes the output of the soft-max layer, $C$ is the number of classes in labels, $y_c(x) \in \{0,1\}$ indicates the binary value at voxel $x$ for each class, $M_c$ indicates the manual segmentation and $PL_{c}$ indicates the predicted label map obtained by determining the argmax of labels from the soft-max output.

\medskip
During testing, the predictions were obtained as 2D probability maps for slices in each plane and were later assembled into 3D volumes and resized to the original dimensions. We then averaged the 3D probability volumes to get the final probability volume ($P_{avg}$) for the triplanar architecture. In addition, we obtained a 3D probability map ($P_{TC}$) from the independent axial U-Net for predicting TC label. Note that 3D probability maps $P_{avg}$ and $P_{TC}$ still have a 4$^{\text{th}}$ dimension corresponding to the labels.

\subsection{Post-processing}
\label{ssec:postproc}
We obtained the predicted ET, NCR/NET and ED label maps by determining the \textit{argmax} of labels (4$^{\text{th}}$ dimension) in $P_{avg}$ and padded them with zeros to bring them back to their original dimensions. Similarly, we obtained the predicted the TC label map from $P_{TC}$ as argmax of TC against background. We then applied the following additional rules, based on prior knowledge and patterns observed in the manual segmentations: predicted ET regions 
with volume $<$ 200 mm$^3$ were relabeled as part of the NCR/NET region; the difference between TC and ET was relabelled as NCR/NET. We also performed a morphological clean-up by removing small isolated noisy stray regions in ED output (volume $<$ 200 mm$^3$ and located at a distance $>$ 75 mm from the centre of the largest ED region) and filled in the missed voxels in the TC - ED interfaces as part of the ED output. 

\subsection{Implementation details}
\label{ssec:implement}

The models were trained using the Adam Optimiser with $\epsilon$=$10^{-4}$. We empirically chose a batch size of 8, with an initial learning rate of $1 \times 10^{-3}$, reducing it by a factor of $1 \times 10^{-1}$ every 2 epochs (set empirically due to the rapid reduction of loss values at these early epochs), until it reached a fixed value of $1 \times 10^{-5}$. Every individual U-Net took $\approx$ 50 epochs for convergence. The networks were trained on an NVIDIA Tesla V100, taking $\approx$ 12 mins (for 3 planes + TC network) per epoch training/validation split of 90\%/10\%.  

\subsection{Data augmentation}
\label{ssec:aug}

Data augmentation was applied in an online manner by randomly selecting from following transformations: translation (x/y-offset $\in$ [-10, 10]), rotation ($\theta$ $\in$ [-10, 10]) and random noise injection (Gaussian, $\mu$ = 0, $\sigma^2$ $\in$ [0.01, 0.09]), increasing the dataset by a factor of 2 (chosen empirically) for all planes. The hyperparameters for the transformations  were randomly sampled from the above specified closed intervals using a uniform distribution. 

\subsection{Performance evaluation metrics}
\label{ssec:perfs}

Metrics computed by the online evaluation platform in BraTS'20 are (i) Dice Similarity Coefficient measured as 2 $\times$ TP / (2 $\times$ TP + FP + FN), (ii) sensitivity measured as TP / (TP + FN), (iii) specificity measured as TN / (TN + FP) and (iv) the 95$^{\text{th}}$ percentile of the Hausdorff Distance (H95), where TP, FP, FN and TN are number of true positive, false positive, false negative and true negative voxels respectively. Regarding the ground truth labels, while the manual segmentations consist of ET, ED and NCR/NET classes, evaluation performance metrics were determined by the evaluation platform for the following sub-regions of tumour: (1) ET, (2) TC (NCR/NET + ET) and (3) whole tumour (WT, which is TC + ED). 

\section{Experiments and results}
\label{sec:exps_results}

\begin{figure*}[h!]
    \centering
    \includegraphics[width=\textwidth]{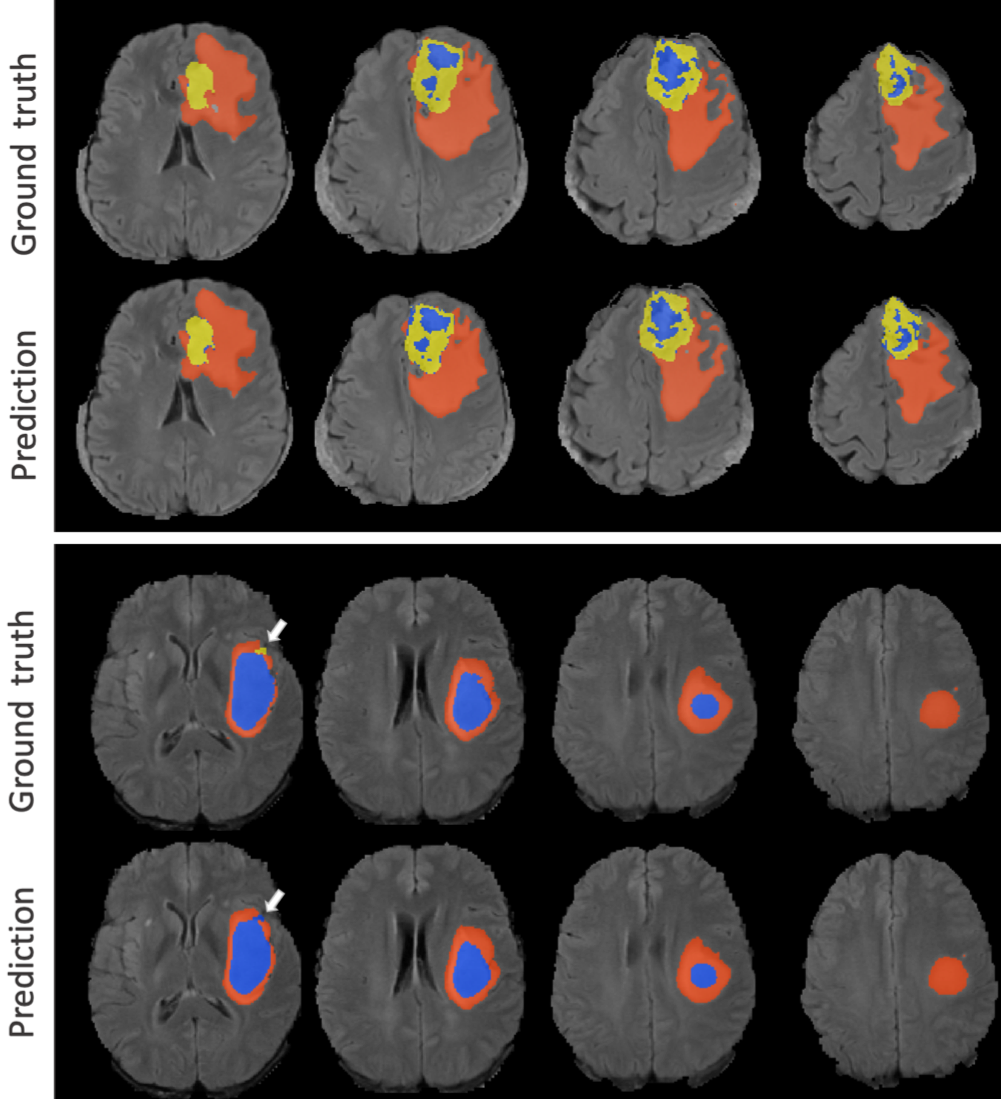}
    \caption[]{\small Results on BraTS'20 training data from two sample subjects (top and bottom panels). Manual segmentations and predicted outputs on axial slices from two sample subjects (NCR/NET - blue, ET - yellow and ED - red). Subject in the top panel: Dice (ET/WT/TC) - 0.96/0.98/0.97, sensitivity (ET/WT/TC) - 0.95/0.98/0.99, specificity (all) - 0.99, H95 (all) - 1 mm; subject in the bottom panel: Dice (ET/WT/TC) - 0/0.98/0.94, sensitivity (ET/WT/TC) - 0/0.97/0.89, specificity (ET/WT/TC) - 1/0.99/1, H95 (ET/WT/TC) - 373.1/1/1.4. In the bottom panel, the white arrows indicate the false prediction of the ET region, leading to the dice value of 0.00. }
    \label{fig:tr_res_vis}
\end{figure*}

\textbf{Cross-validation on the labelled training data: }We used 369 labelled subjects from BraTS'20 training data to perform 5-fold cross-validation with a training-validation-testing split ratio of 255-37-73 subjects (255-37-77 for the last fold). The results on the test splits (evaluated using the challenge online platform) are shown in table~\ref{table:tr_res}. Among the three sub-regions, we achieved the best segmentation performance for the whole tumour (WT) with a mean Dice value of 0.93. A few sample outputs of our method are shown along with manual segmentations in fig.~\ref{fig:tr_res_vis}.

\begin{table*}[h!]
\centering
\scriptsize
\caption{Results of 5-fold cross-validation on BraTS'20 Training data.}
{\renewcommand{\arraystretch}{1.5}%
\begin{tabular}{l|ccc|ccc|ccc|ccc}%|>{\setlength{\baselineskip}{1.5\baselineskip}}c|}
  \hline
  \parbox[c][][t]{1.6cm}{\raggedright }& 
  \multicolumn{3}{|c|}{\parbox[c][][t]{2.4cm}{\centering \textbf{Dice}}}& 
  \multicolumn{3}{|c|}{\parbox[c][][t]{2.4cm}{\centering \textbf{Sensitivity}}} & 
  \multicolumn{3}{|c|}{\parbox[c][][t]{2.4cm}{\centering \textbf{Specificity}}} & 
  \multicolumn{3}{|c}{\parbox[c][][t]{2.1cm}{\centering \textbf{H95 (mm)}}} \\
  \cline{2-13}
  \parbox[c][][t]{1.6cm}{\raggedright }& 
  \parbox[c][][t]{0.8cm}{\centering \textbf{ET}} &
  \parbox[c][][t]{0.8cm}{\centering \textbf{WT}} &
  \parbox[c][][t]{0.8cm}{\centering \textbf{TC}} &
  \parbox[c][][t]{0.8cm}{\centering \textbf{ET}} &
  \parbox[c][][t]{0.8cm}{\centering \textbf{WT}} &
  \parbox[c][][t]{0.8cm}{\centering \textbf{TC}} &
  \parbox[c][][t]{0.8cm}{\centering \textbf{ET}} &
  \parbox[c][][t]{0.8cm}{\centering \textbf{WT}} &
  \parbox[c][][t]{0.8cm}{\centering \textbf{TC}} &
  \parbox[c][][t]{0.7cm}{\centering \textbf{ET}} &
  \parbox[c][][t]{0.7cm}{\centering \textbf{WT}} &
  \parbox[c][][t]{0.7cm}{\centering \textbf{TC}} \\
  \hline
  \parbox[c][][t]{1.6cm}{\raggedright \textbf{Mean}}& 
  \parbox[c][][t]{0.8cm}{\centering 0.83} &
  \parbox[c][][t]{0.8cm}{\centering 0.93} &
  \parbox[c][][t]{0.8cm}{\centering 0.87} &
  \parbox[c][][t]{0.8cm}{\centering 0.83} &
  \parbox[c][][t]{0.8cm}{\centering 0.90} &
  \parbox[c][][t]{0.8cm}{\centering 0.85} &
  \parbox[c][][t]{0.8cm}{\centering 0.99} &
  \parbox[c][][t]{0.8cm}{\centering 0.99} &
  \parbox[c][][t]{0.8cm}{\centering 0.99} &
  \parbox[c][][t]{0.7cm}{\centering 20.3} &
  \parbox[c][][t]{0.7cm}{\centering 3.4} &
  \parbox[c][][t]{0.7cm}{\centering 6.3} \\
  \hline
  \parbox[c][][t]{1.6cm}{\raggedright \textbf{Std.}}& 
  \parbox[c][][t]{0.8cm}{\centering {0.22}} &
  \parbox[c][][t]{0.8cm}{\centering {0.08}} &
  \parbox[c][][t]{0.8cm}{\centering {0.17}} &
  \parbox[c][][t]{0.8cm}{\centering {0.23}} &
  \parbox[c][][t]{0.8cm}{\centering {0.11}} &
  \parbox[c][][t]{0.8cm}{\centering {0.19}} &
  \parbox[c][][t]{0.8cm}{\centering {.0005}} &
  \parbox[c][][t]{0.8cm}{\centering {.0005}} &
  \parbox[c][][t]{0.8cm}{\centering {.0005}} &
  \parbox[c][][t]{0.7cm}{\centering {80.3}} &
  \parbox[c][][t]{0.7cm}{\centering {5.1}} &
  \parbox[c][][t]{0.7cm}{\centering {28.0}} \\
  \hline
  \parbox[c][][t]{1.6cm}{\raggedright \textbf{Median}}& 
  \parbox[c][][t]{0.8cm}{\centering {0.90}} &
  \parbox[c][][t]{0.8cm}{\centering {0.95}} &
  \parbox[c][][t]{0.8cm}{\centering {0.93}} &
  \parbox[c][][t]{0.8cm}{\centering {0.91}} &
  \parbox[c][][t]{0.8cm}{\centering {0.93}} &
  \parbox[c][][t]{0.8cm}{\centering {0.93}} &
  \parbox[c][][t]{0.8cm}{\centering {0.99}} &
  \parbox[c][][t]{0.8cm}{\centering {0.99}} &
  \parbox[c][][t]{0.8cm}{\centering {0.99}} &
  \parbox[c][][t]{0.7cm}{\centering {1.0}} &
  \parbox[c][][t]{0.7cm}{\centering {1.7}} &
  \parbox[c][][t]{0.7cm}{\centering {2.2}} \\
  \hline
  \parbox[c][][t]{1.6cm}{\raggedright \textbf{25 quantile}}& 
  \parbox[c][][t]{0.8cm}{\centering {0.83}} &
  \parbox[c][][t]{0.8cm}{\centering {0.91}} &
  \parbox[c][][t]{0.8cm}{\centering {0.86}} &
  \parbox[c][][t]{0.8cm}{\centering {0.83}} &
  \parbox[c][][t]{0.8cm}{\centering {0.87}} &
  \parbox[c][][t]{0.8cm}{\centering {0.83}} &
  \parbox[c][][t]{0.8cm}{\centering {0.99}} &
  \parbox[c][][t]{0.8cm}{\centering {0.99}} &
  \parbox[c][][t]{0.8cm}{\centering {0.99}} &
  \parbox[c][][t]{0.7cm}{\centering {1.0}} &
  \parbox[c][][t]{0.7cm}{\centering {1.0}} &
  \parbox[c][][t]{0.7cm}{\centering {1.4}} \\
  \hline
  \parbox[c][][t]{1.6cm}{\raggedright \textbf{75 quantile}}& 
  \parbox[c][][t]{0.8cm}{\centering {0.95}} &
  \parbox[c][][t]{0.8cm}{\centering {0.97}} &
  \parbox[c][][t]{0.8cm}{\centering {0.96}} &
  \parbox[c][][t]{0.8cm}{\centering {0.96}} &
  \parbox[c][][t]{0.8cm}{\centering {0.97}} &
  \parbox[c][][t]{0.8cm}{\centering {0.97}} &
  \parbox[c][][t]{0.8cm}{\centering {0.99}} &
  \parbox[c][][t]{0.8cm}{\centering {0.99}} &
  \parbox[c][][t]{0.8cm}{\centering {0.99}} &
  \parbox[c][][t]{0.7cm}{\centering {2.2}} &
  \parbox[c][][t]{0.7cm}{\centering {3.2}} &
  \parbox[c][][t]{0.7cm}{\centering {4.2}} \\
  \hline
  \multicolumn{13}{c}{\parbox[c][][t]{10.9cm}{\centering ET - enhancing tumour, WT - whole tumour, TC - tumour core.}}\\
  \end{tabular}}
% \vspace{-1.5em}
\label{table:tr_res}
\end{table*}

\begin{figure*}[h!]
    \centering
    \includegraphics[width=\textwidth]{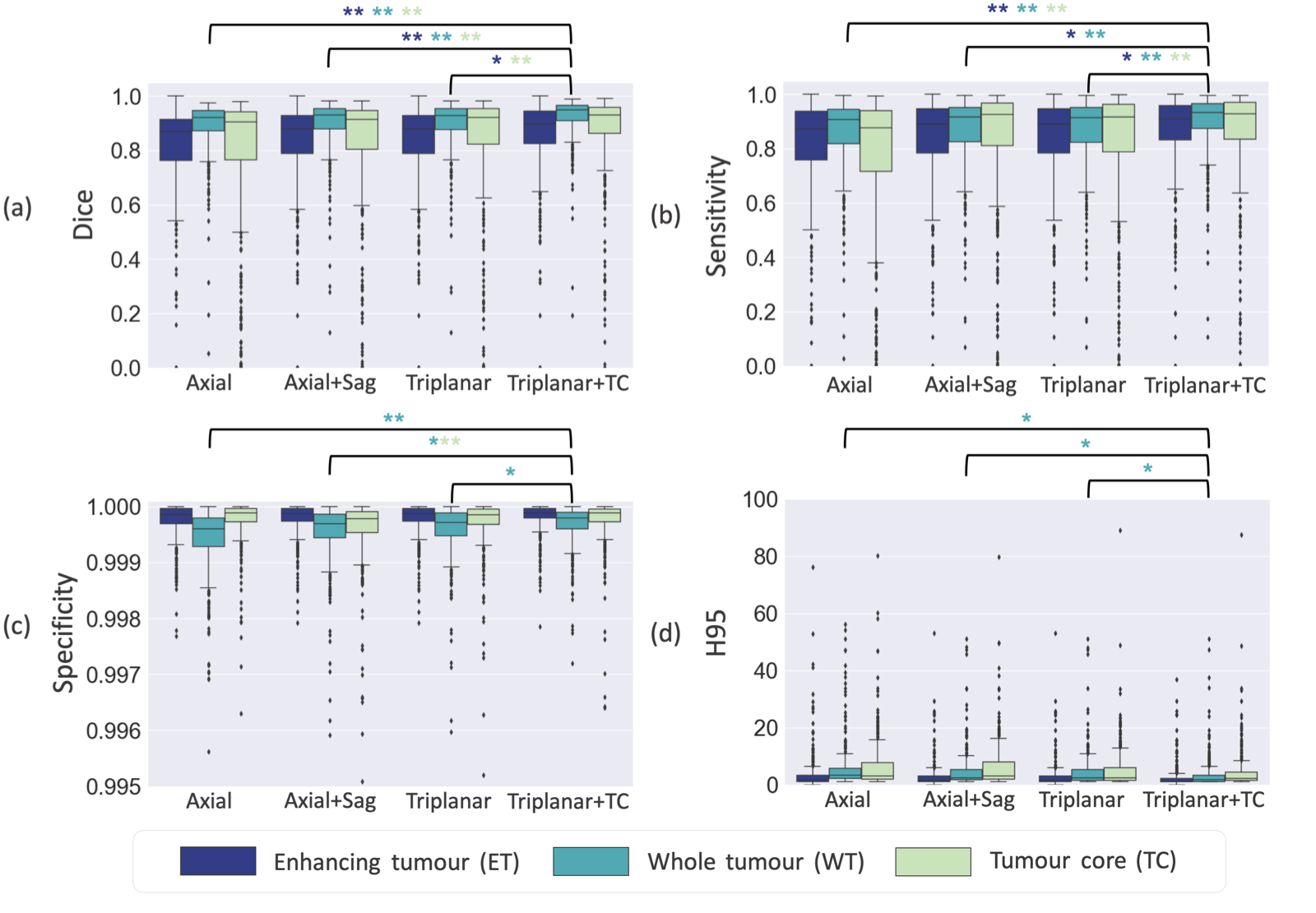}
    \caption[]{Boxplots of results from the ablation study, showing (a) Dice, (b) sensitivity, (c) specificity and (d) H95 for axial, axial+sagittal, triplanar and triplanar+TC cases. Significant differences in performance metrics are indicated by asterisks (* - p $<$ 0.01, ** - p $<$ 0.001) in the corresponding colours for tumour sub-regions. Note that only the significant differences in performance metrics between triplanar+TC and other cases are shown (for all combinations of individual pairs, refer to table~\ref{table:abl_study_res}).}
    \label{fig:abl_study_res}
\end{figure*}

\medskip
\hspace{-0.6cm} \textbf{Ablation study on the labelled training data: }In order to determine the effect of individual components of our architecture on the segmentation performance, we evaluated the segmentation results with the following components: (i) axial 2D U-Net only, (ii) axial + sagittal 2D U-Nets, (iii) triplanar network (axial + sagittal + coronal U-Nets) and (iv) triplanar network + axial U-Net for TC label detection. We used a cross-validation strategy (with the same training-validation-test split mentioned above) to evaluate the performance metrics. The values of performance metrics for the ablation study are shown in table~\ref{table:abl_study_res} and the corresponding boxplots are shown in fig.~\ref{fig:abl_study_res}. The segmentation of sub-regions improved with the addition of the TC network, with significant increase in Dice and sensitivity values (p $<$ 0.01), especially for the ET and TC sub-regions. We also observed significant improvement in the specificity of the WT segmentation (p $<$ 0.001) using the triplanar architecture when compared to individual U-Nets.

\begin{table*}[h!]
\centering
\scriptsize
\caption{Ablation study results for BraTS'20 Training data. Mean and standard deviation (in brackets) values for various components of the method reported for tumour sub-regions. In the lower part p-values of paired two-tailed t-test results between individual pairs of components are shown, with significant values underlined.}
{\renewcommand{\arraystretch}{1.5}%
\begin{tabular}{l|ccc|ccc|ccc|ccc}%|>{\setlength{\baselineskip}{1.5\baselineskip}}c|}
  \hline
  \parbox[c][][t]{1.6cm}{\raggedright }& 
  \multicolumn{3}{|c|}{\parbox[c][][t]{2.4cm}{\centering \textbf{Dice}}}& 
  \multicolumn{3}{|c|}{\parbox[c][][t]{2.4cm}{\centering \textbf{Sensitivity}}} & 
  \multicolumn{3}{|c|}{\parbox[c][][t]{2.4cm}{\centering \textbf{Specificity}}} & 
  \multicolumn{3}{|c}{\parbox[c][][t]{2.1cm}{\centering \textbf{H95 (mm)}}} \\
  \cline{2-13}
  \parbox[c][][t]{1.6cm}{\raggedright }& 
  \parbox[c][][t]{0.8cm}{\centering \textbf{ET}} &
  \parbox[c][][t]{0.8cm}{\centering \textbf{WT}} &
  \parbox[c][][t]{0.8cm}{\centering \textbf{TC}} &
  \parbox[c][][t]{0.8cm}{\centering \textbf{ET}} &
  \parbox[c][][t]{0.8cm}{\centering \textbf{WT}} &
  \parbox[c][][t]{0.8cm}{\centering \textbf{TC}} &
  \parbox[c][][t]{0.8cm}{\centering \textbf{ET}} &
  \parbox[c][][t]{0.8cm}{\centering \textbf{WT}} &
  \parbox[c][][t]{0.8cm}{\centering \textbf{TC}} &
  \parbox[c][][t]{0.7cm}{\centering \textbf{ET}} &
  \parbox[c][][t]{0.7cm}{\centering \textbf{WT}} &
  \parbox[c][][t]{0.7cm}{\centering \textbf{TC}} \\
  \hline
  \parbox[c][][t]{1.6cm}{\raggedright \textbf{A}}& 
  \parbox[c][][t]{0.8cm}{\vspace{0.35em} \centering {0.78\\(0.25)}} &
  \parbox[c][][t]{0.8cm}{\vspace{0.35em} \centering {0.89\\(0.10)}} &
  \parbox[c][][t]{0.8cm}{\vspace{0.35em} \centering {0.80\\(0.24)}} &
  \parbox[c][][t]{0.8cm}{\vspace{0.35em} \centering {0.78\\(0.26)}} &
  \parbox[c][][t]{0.8cm}{\vspace{0.35em} \centering {0.86\\(0.13)}} &
  \parbox[c][][t]{0.8cm}{\vspace{0.35em} \centering {0.76\\(0.26)}} &
  \parbox[c][][t]{0.8cm}{\vspace{0.35em} \centering {0.9997\\\tiny{(.0006)}}} &
  \parbox[c][][t]{0.8cm}{\vspace{0.35em} \centering {0.9994\\\tiny{(.0007)}}} &
  \parbox[c][][t]{0.8cm}{\vspace{0.35em} \centering {0.9997\\\tiny{(.0005)}}} &
  \parbox[c][][t]{0.7cm}{\vspace{0.35em} \centering {26.6\\(89.7)}} &
  \parbox[c][][t]{0.7cm}{\vspace{0.35em} \centering {5.7\\(7.9)}} &
  \parbox[c][][t]{0.7cm}{\vspace{0.35em} \centering {7.3\\(20.9)}} \\
  \parbox[c][][t]{1.6cm}{\raggedright \textbf{A+S}}& 
  \parbox[c][][t]{0.8cm}{\vspace{0.35em} \centering {0.79\\(0.24)}} &
  \parbox[c][][t]{0.8cm}{\vspace{0.35em} \centering {0.90\\(0.10)}} &
  \parbox[c][][t]{0.8cm}{\vspace{0.35em} \centering {0.82\\(0.21)}} &
  \parbox[c][][t]{0.8cm}{\vspace{0.35em} \centering {0.79\\(0.26)}} &
  \parbox[c][][t]{0.8cm}{\vspace{0.35em} \centering {0.87\\(0.13)}} &
  \parbox[c][][t]{0.8cm}{\vspace{0.35em} \centering {0.83\\(0.23)}} &
  \parbox[c][][t]{0.8cm}{\vspace{0.35em} \centering {0.9997\\\tiny{(.0006)}}} &
  \parbox[c][][t]{0.8cm}{\vspace{0.35em} \centering {0.9995\\\tiny{(.0006)}}} &
  \parbox[c][][t]{0.8cm}{\vspace{0.35em} \centering {0.9995\\\tiny{(.0008)}}} &
  \parbox[c][][t]{0.7cm}{\vspace{0.35em} \centering {25.9\\(89.7)}} &
  \parbox[c][][t]{0.7cm}{\vspace{0.35em} \centering {4.9\\(8.4)}} &
  \parbox[c][][t]{0.7cm}{\vspace{0.35em} \centering {7.4\\(21.4)}} \\
  \parbox[c][][t]{1.6cm}{\raggedright \textbf{TP}}& 
  \parbox[c][][t]{0.8cm}{\vspace{0.35em} \centering {0.79\\(0.24)}} &
  \parbox[c][][t]{0.8cm}{\vspace{0.35em} \centering {0.90\\(0.09)}} &
  \parbox[c][][t]{0.8cm}{\vspace{0.35em} \centering {0.83\\(0.21)}} &
  \parbox[c][][t]{0.8cm}{\vspace{0.35em} \centering {0.79\\(0.26)}} &
  \parbox[c][][t]{0.8cm}{\vspace{0.35em} \centering {0.86\\(0.14)}} &
  \parbox[c][][t]{0.8cm}{\vspace{0.35em} \centering {0.82\\(0.23)}} &
  \parbox[c][][t]{0.8cm}{\vspace{0.35em} \centering {0.9997\\\tiny{(.0006)}}} &
  \parbox[c][][t]{0.8cm}{\vspace{0.35em} \centering {0.9996\\\tiny{(.0006)}}} &
  \parbox[c][][t]{0.8cm}{\vspace{0.35em} \centering {0.9997\\\tiny{(.0006)}}} &
  \parbox[c][][t]{0.7cm}{\vspace{0.35em} \centering {25.8\\(89.6)}} &
  \parbox[c][][t]{0.7cm}{\vspace{0.35em} \centering {4.6\\(6.1)}} &
  \parbox[c][][t]{0.7cm}{\vspace{0.35em} \centering {7.0\\(28.0)}} \\
  \parbox[c][][t]{1.6cm}{\raggedright \textbf{TP+TC}}& 
  \parbox[c][][t]{0.8cm}{\vspace{0.35em} \centering {0.82\\(0.22)}} &
  \parbox[c][][t]{0.8cm}{\vspace{0.35em} \centering {0.92\\(0.07)}} &
  \parbox[c][][t]{0.8cm}{\vspace{0.35em} \centering {0.87\\(0.17)}} &
  \parbox[c][][t]{0.8cm}{\vspace{0.35em} \centering {0.83\\(0.23)}} &
  \parbox[c][][t]{0.8cm}{\vspace{0.35em} \centering {0.90\\(0.11)}} &
  \parbox[c][][t]{0.8cm}{\vspace{0.35em} \centering {0.85\\(0.20)}} &
  \parbox[c][][t]{0.8cm}{\vspace{0.35em} \centering {0.9998\\\tiny{(.0005)}}} &
  \parbox[c][][t]{0.8cm}{\vspace{0.35em} \centering {0.9997\\\tiny{(.0005)}}} &
  \parbox[c][][t]{0.8cm}{\vspace{0.35em} \centering {0.9997\\\tiny{(.0005)}}} &
  \parbox[c][][t]{0.7cm}{\vspace{0.35em} \centering {20.3\\(80.1)}} &
  \parbox[c][][t]{0.7cm}{\vspace{0.35em} \centering {3.4\\(5.3)}} &
  \parbox[c][][t]{0.7cm}{\vspace{0.35em} \centering {6.3\\(28.0)}} \\
  \hline
  \multicolumn{13}{c}{\parbox[c][][t]{10cm}{\centering \textbf{p-values}}} \\
  \hline
  \parbox[c][][t]{1.6cm}{\raggedright \textbf{A vs A+S}}& 
  \parbox[c][][t]{0.8cm}{\centering {0.36}} &
  \parbox[c][][t]{0.8cm}{\centering {0.20}} &
  \parbox[c][][t]{0.8cm}{\centering {0.10}} &
  \parbox[c][][t]{0.8cm}{\centering {0.48}} &
  \parbox[c][][t]{0.8cm}{\centering {0.60}} &
  \parbox[c][][t]{0.8cm}{\centering \tiny \underline{\textless 0.001}} &
  \parbox[c][][t]{0.8cm}{\centering {0.43}} &
  \parbox[c][][t]{0.8cm}{\centering \underline{0.006}} &
  \parbox[c][][t]{0.8cm}{\centering \tiny \underline{\textless 0.001}} &
  \parbox[c][][t]{0.7cm}{\centering {0.90}} &
  \parbox[c][][t]{0.7cm}{\centering {0.22}} &
  \parbox[c][][t]{0.7cm}{\centering {0.94}} \\
  \parbox[c][][t]{1.6cm}{\raggedright \textbf{A vs TP}}& 
  \parbox[c][][t]{0.8cm}{\centering {0.37}} &
  \parbox[c][][t]{0.8cm}{\centering {0.17}} &
  \parbox[c][][t]{0.8cm}{\centering \underline{0.03}} &
  \parbox[c][][t]{0.8cm}{\centering {0.47}} &
  \parbox[c][][t]{0.8cm}{\centering {0.77}} &
  \parbox[c][][t]{0.8cm}{\centering \underline{0.004}} &
  \parbox[c][][t]{0.8cm}{\centering {0.42}} &
  \parbox[c][][t]{0.8cm}{\centering \tiny \underline{\textless 0.001}} &
  \parbox[c][][t]{0.8cm}{\centering {0.19}} &
  \parbox[c][][t]{0.7cm}{\centering {0.90}} &
  \parbox[c][][t]{0.7cm}{\centering \underline{0.03}} &
  \parbox[c][][t]{0.7cm}{\centering {0.92}} \\
  \parbox[c][][t]{1.6cm}{\raggedright \textbf{A vs TP+TC}}& 
  \parbox[c][][t]{0.8cm}{\centering \underline{0.005}} &
  \parbox[c][][t]{0.8cm}{\centering \tiny \underline{\textless 0.001}} &
  \parbox[c][][t]{0.8cm}{\centering \tiny \underline{\textless 0.001}} &
  \parbox[c][][t]{0.8cm}{\centering \underline{0.007}} &
  \parbox[c][][t]{0.8cm}{\centering \tiny \underline{\textless 0.001}} &
  \parbox[c][][t]{0.8cm}{\centering \tiny \underline{\textless 0.001}} &
  \parbox[c][][t]{0.8cm}{\centering {0.14}} &
  \parbox[c][][t]{0.8cm}{\centering \tiny \underline{\textless 0.001}} &
  \parbox[c][][t]{0.8cm}{\centering {0.77}} &
  \parbox[c][][t]{0.7cm}{\centering {0.31}} &
  \parbox[c][][t]{0.7cm}{\centering \tiny \underline{\textless 0.001}} &
  \parbox[c][][t]{0.7cm}{\centering {0.60}} \\
  \hline
  \parbox[c][][t]{1.6cm}{\raggedright \textbf{A+S vs TP}}& 
  \parbox[c][][t]{0.8cm}{\centering {0.99}} &
  \parbox[c][][t]{0.8cm}{\centering {0.93}} &
  \parbox[c][][t]{0.8cm}{\centering {0.56}} &
  \parbox[c][][t]{0.8cm}{\centering {0.99}} &
  \parbox[c][][t]{0.8cm}{\centering {0.82}} &
  \parbox[c][][t]{0.8cm}{\centering {0.60}} &
  \parbox[c][][t]{0.8cm}{\centering {0.99}} &
  \parbox[c][][t]{0.8cm}{\centering {0.40}} &
  \parbox[c][][t]{0.8cm}{\centering \underline{0.01}} &
  \parbox[c][][t]{0.7cm}{\centering {0.99}} &
  \parbox[c][][t]{0.7cm}{\centering {0.48}} &
  \parbox[c][][t]{0.7cm}{\centering {0.87}} \\
  \parbox[c][][t]{1.6cm}{\raggedright \textbf{A+S vs TP+TC}}& 
  \parbox[c][][t]{0.8cm}{\centering {0.06}} &
  \parbox[c][][t]{0.8cm}{\centering \tiny \underline{\textless 0.001}} &
  \parbox[c][][t]{0.8cm}{\centering \underline{0.002}} &
  \parbox[c][][t]{0.8cm}{\centering \underline{0.04}} &
  \parbox[c][][t]{0.8cm}{\centering \tiny \underline{\textless 0.001}} &
  \parbox[c][][t]{0.8cm}{\centering {0.08}} &
  \parbox[c][][t]{0.8cm}{\centering {0.50}} &
  \parbox[c][][t]{0.8cm}{\centering \underline{0.001}} &
  \parbox[c][][t]{0.8cm}{\centering \tiny \underline{\textless 0.001}} &
  \parbox[c][][t]{0.7cm}{\centering {0.38}} &
  \parbox[c][][t]{0.7cm}{\centering \underline{0.002}} &
  \parbox[c][][t]{0.7cm}{\centering {0.55}} \\
  \hline
  \parbox[c][][t]{1.6cm}{\vspace{0.2em} \raggedright \textbf{TP vs TP+TC}}& 
  \parbox[c][][t]{0.8cm}{\centering {0.06}} &
  \parbox[c][][t]{0.8cm}{\centering \tiny \underline{\textless 0.001}} &
  \parbox[c][][t]{0.8cm}{\centering \underline{0.01}} &
  \parbox[c][][t]{0.8cm}{\centering \underline{0.04}} &
  \parbox[c][][t]{0.8cm}{\centering \tiny \underline{\textless 0.001}} &
  \parbox[c][][t]{0.8cm}{\centering \underline{0.02}} &
  \parbox[c][][t]{0.8cm}{\centering {0.50}} &
  \parbox[c][][t]{0.8cm}{\centering \underline{0.02}} &
  \parbox[c][][t]{0.8cm}{\centering {0.30}} &
  \parbox[c][][t]{0.7cm}{\centering {0.38}} &
  \parbox[c][][t]{0.7cm}{\centering \underline{0.004}} &
  \parbox[c][][t]{0.7cm}{\centering {0.70}} \\
  \hline
  \multicolumn{13}{c}{\parbox[c][][t]{11.9cm}{\vspace{0.3em} \centering Tumour sub-regions: ET - enhancing tumour, WT - whole tumour, TC - tumour core. \\Methods: A - axial, A+S - axial + sagittal, TP - triplanar, TP+TC - triplanar + TC network.}}\\
  \end{tabular}}
% \vspace{-1.5em}
\label{table:abl_study_res}
\end{table*}

\begin{figure*}[h!]
    \centering
    \includegraphics[width=10.5cm]{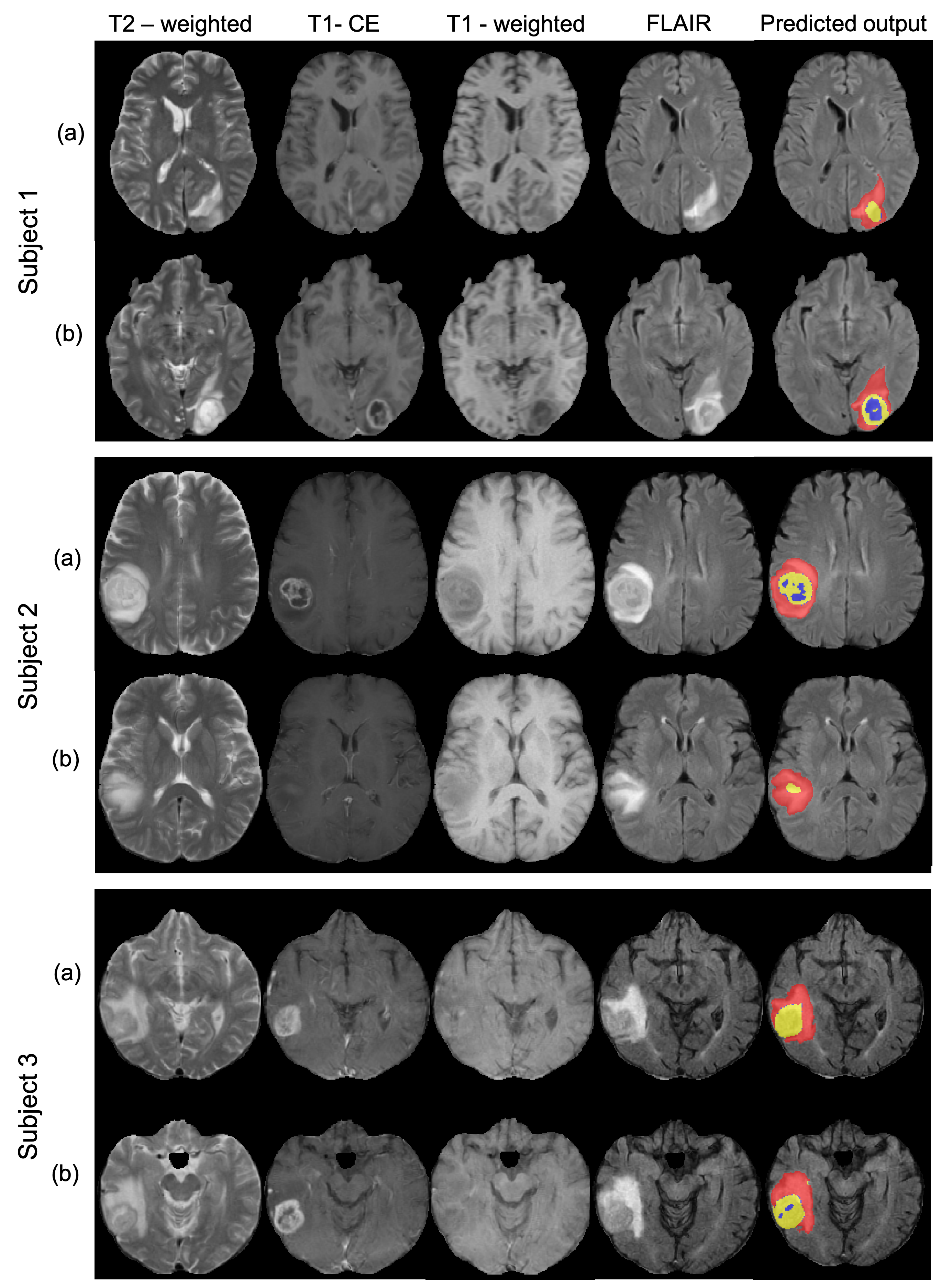}
    \caption[]{\small Results on BraTS'20 validation data from three sample subjects. Predicted outputs on two axial slices (a) and (b) from from three sample subjects (NCR/NET - blue, ET - yellow and ED - red), along with the provided input modalities: T2-weighted, T1-CE, T1 and FLAIR. Subject1: Dice (ET/WT/TC) - 0.86/0.97/0.96, sensitivity (ET/WT/TC) - 0.81/0.96/0.98, specificity (all) - 0.99, H95 (ET/WT/TC) - 1.4/1/1 mm; Subject2: Dice (ET/WT/TC) - 0.93/0.97/0.96, sensitivity (ET/WT/TC) - 0.93/0.99/0.97, specificity (all) - 0.99, H95 (all) - 1 mm; Subject3: Dice (ET/WT/TC) - 0.95/0.93/0.95, sensitivity (ET/WT/TC) - 0.97/0.99/0.98, specificity (all) - 0.99, H95 (ET/WT/TC) - 1/2/1 mm.}
    \label{fig:val_res_vis}
\end{figure*}

\medskip
\hspace{-0.6cm} \textbf{Evaluation on the unlabelled validation data: } The models trained from the 5-fold cross-validation of the training data were applied to the validation data and the final predictions were obtained using majority voting (from 5 models). The results obtained from the online evaluation platform are shown in table~\ref{table:val_res}. A few sample validation data outputs of our method are shown along with input modalities in fig.~\ref{fig:val_res_vis}. The results followed a trend similar to the training data obtaining the best results for WT segmentation with mean Dice and H95 values of 0.89 and 4.4 respectively.

\begin{table*}[h!]
\centering
\scriptsize
\caption{Results for BraTS'20 Validation data.}
{\renewcommand{\arraystretch}{1.5}%
\begin{tabular}{l|ccc|ccc|ccc|ccc}%|>{\setlength{\baselineskip}{1.5\baselineskip}}c|}
  \hline
  \parbox[c][][t]{1.6cm}{\raggedright }& 
  \multicolumn{3}{|c|}{\parbox[c][][t]{2.4cm}{\centering \textbf{Dice}}}& 
  \multicolumn{3}{|c|}{\parbox[c][][t]{2.4cm}{\centering \textbf{Sensitivity}}} & 
  \multicolumn{3}{|c|}{\parbox[c][][t]{2.4cm}{\centering \textbf{Specificity}}} & 
  \multicolumn{3}{|c}{\parbox[c][][t]{2.1cm}{\centering \textbf{H95 (mm)}}} \\
  \cline{2-13}
  \parbox[c][][t]{1.6cm}{\raggedright }& 
  \parbox[c][][t]{0.8cm}{\centering \textbf{ET}} &
  \parbox[c][][t]{0.8cm}{\centering \textbf{WT}} &
  \parbox[c][][t]{0.8cm}{\centering \textbf{TC}} &
  \parbox[c][][t]{0.8cm}{\centering \textbf{ET}} &
  \parbox[c][][t]{0.8cm}{\centering \textbf{WT}} &
  \parbox[c][][t]{0.8cm}{\centering \textbf{TC}} &
  \parbox[c][][t]{0.8cm}{\centering \textbf{ET}} &
  \parbox[c][][t]{0.8cm}{\centering \textbf{WT}} &
  \parbox[c][][t]{0.8cm}{\centering \textbf{TC}} &
  \parbox[c][][t]{0.7cm}{\centering \textbf{ET}} &
  \parbox[c][][t]{0.7cm}{\centering \textbf{WT}} &
  \parbox[c][][t]{0.7cm}{\centering \textbf{TC}} \\
  \hline
  \parbox[c][][t]{1.6cm}{\raggedright \textbf{Mean}}& 
  \parbox[c][][t]{0.8cm}{\centering {0.77}} &
  \parbox[c][][t]{0.8cm}{\centering {0.89}} &
  \parbox[c][][t]{0.8cm}{\centering {0.77}} &
  \parbox[c][][t]{0.8cm}{\centering {0.76}} &
  \parbox[c][][t]{0.8cm}{\centering {0.87}} &
  \parbox[c][][t]{0.8cm}{\centering {0.73}} &
  \parbox[c][][t]{0.8cm}{\centering {0.99}} &
  \parbox[c][][t]{0.8cm}{\centering {0.99}} &
  \parbox[c][][t]{0.8cm}{\centering {0.99}} &
  \parbox[c][][t]{0.7cm}{\centering {29.4}} &
  \parbox[c][][t]{0.7cm}{\centering {4.4}} &
  \parbox[c][][t]{0.7cm}{\centering {15.3}} \\
  \hline
  \parbox[c][][t]{1.6cm}{\raggedright \textbf{Std.}}& 
  \parbox[c][][t]{0.8cm}{\centering {0.27}} &
  \parbox[c][][t]{0.8cm}{\centering {0.09}} &
  \parbox[c][][t]{0.8cm}{\centering {0.27}} &
  \parbox[c][][t]{0.8cm}{\centering {0.29}} &
  \parbox[c][][t]{0.8cm}{\centering {0.13}} &
  \parbox[c][][t]{0.8cm}{\centering {0.29}} &
  \parbox[c][][t]{0.8cm}{\centering {.0005}} &
  \parbox[c][][t]{0.8cm}{\centering {.0009}} &
  \parbox[c][][t]{0.8cm}{\centering {.0003}} &
  \parbox[c][][t]{0.7cm}{\centering {96.2}} &
  \parbox[c][][t]{0.7cm}{\centering {5.4}} &
  \parbox[c][][t]{0.7cm}{\centering {57.3}} \\
  \hline
  \parbox[c][][t]{1.6cm}{\raggedright \textbf{Median}}& 
  \parbox[c][][t]{0.8cm}{\centering {0.87}} &
  \parbox[c][][t]{0.8cm}{\centering {0.93}} &
  \parbox[c][][t]{0.8cm}{\centering {0.90}} &
  \parbox[c][][t]{0.8cm}{\centering {0.88}} &
  \parbox[c][][t]{0.8cm}{\centering {0.91}} &
  \parbox[c][][t]{0.8cm}{\centering {0.86}} &
  \parbox[c][][t]{0.8cm}{\centering {0.99}} &
  \parbox[c][][t]{0.8cm}{\centering {0.99}} &
  \parbox[c][][t]{0.8cm}{\centering {0.99}} &
  \parbox[c][][t]{0.7cm}{\centering {2.0}} &
  \parbox[c][][t]{0.7cm}{\centering {3.0}} &
  \parbox[c][][t]{0.7cm}{\centering {3.0}} \\
  \hline
  \parbox[c][][t]{1.6cm}{\raggedright \textbf{25 quantile}}& 
  \parbox[c][][t]{0.8cm}{\centering {0.77}} &
  \parbox[c][][t]{0.8cm}{\centering {0.89}} &
  \parbox[c][][t]{0.8cm}{\centering {0.73}} &
  \parbox[c][][t]{0.8cm}{\centering {0.74}} &
  \parbox[c][][t]{0.8cm}{\centering {0.85}} &
  \parbox[c][][t]{0.8cm}{\centering {0.61}} &
  \parbox[c][][t]{0.8cm}{\centering {0.99}} &
  \parbox[c][][t]{0.8cm}{\centering {0.99}} &
  \parbox[c][][t]{0.8cm}{\centering {0.99}} &
  \parbox[c][][t]{0.7cm}{\centering {1.0}} &
  \parbox[c][][t]{0.7cm}{\centering {2.0}} &
  \parbox[c][][t]{0.7cm}{\centering {1.7}} \\
  \hline
  \parbox[c][][t]{1.6cm}{\raggedright \textbf{75 quantile}}& 
  \parbox[c][][t]{0.8cm}{\centering {0.91}} &
  \parbox[c][][t]{0.8cm}{\centering {0.95}} &
  \parbox[c][][t]{0.8cm}{\centering {0.94}} &
  \parbox[c][][t]{0.8cm}{\centering {0.94}} &
  \parbox[c][][t]{0.8cm}{\centering {0.95}} &
  \parbox[c][][t]{0.8cm}{\centering {0.93}} &
  \parbox[c][][t]{0.8cm}{\centering {0.99}} &
  \parbox[c][][t]{0.8cm}{\centering {0.99}} &
  \parbox[c][][t]{0.8cm}{\centering {0.99}} &
  \parbox[c][][t]{0.7cm}{\centering {3.6}} &
  \parbox[c][][t]{0.7cm}{\centering {4.6}} &
  \parbox[c][][t]{0.7cm}{\centering {8.5}} \\
  \hline
  \multicolumn{13}{c}{\parbox[c][][t]{10.9cm}{\centering ET - enhancing tumour, WT - whole tumour, TC - tumour core.}}\\
  \end{tabular}}
% \vspace{-1.5em}
\label{table:val_res}
\end{table*}

\medskip
\hspace{-0.6cm} \textbf{Results on the unseen test data: }Finally, the results obtained from the online evaluation platform on the BraTS'20 test data are shown in table~\ref{table:test_res}. Similar to the validation stage, majority voting on the results of 5 models (from 5-fold cross-validation) was used to predict the tumour region labels. As seen from the table, the Dice values for the WT region was higher than for the ET and TC regions as in the case of the validation stage. However, the Dice and sensitivity values for the ET and TC regions were considerably higher than those on the validation data, and were almost on par with those obtained with the fold-validation on the training dataset.

\begin{table*}[h!]
\centering
\scriptsize
\caption{Results for BraTS'20 Test data.}
{\renewcommand{\arraystretch}{1.5}%
\begin{tabular}{l|ccc|ccc|ccc|ccc}%|>{\setlength{\baselineskip}{1.5\baselineskip}}c|}
  \hline
  \parbox[c][][t]{1.6cm}{\raggedright }& 
  \multicolumn{3}{|c|}{\parbox[c][][t]{2.4cm}{\centering \textbf{Dice}}}& 
  \multicolumn{3}{|c|}{\parbox[c][][t]{2.4cm}{\centering \textbf{Sensitivity}}} & 
  \multicolumn{3}{|c|}{\parbox[c][][t]{2.4cm}{\centering \textbf{Specificity}}} & 
  \multicolumn{3}{|c}{\parbox[c][][t]{2.1cm}{\centering \textbf{H95 (mm)}}} \\
  \cline{2-13}
  \parbox[c][][t]{1.6cm}{\raggedright }& 
  \parbox[c][][t]{0.8cm}{\centering \textbf{ET}} &
  \parbox[c][][t]{0.8cm}{\centering \textbf{WT}} &
  \parbox[c][][t]{0.8cm}{\centering \textbf{TC}} &
  \parbox[c][][t]{0.8cm}{\centering \textbf{ET}} &
  \parbox[c][][t]{0.8cm}{\centering \textbf{WT}} &
  \parbox[c][][t]{0.8cm}{\centering \textbf{TC}} &
  \parbox[c][][t]{0.8cm}{\centering \textbf{ET}} &
  \parbox[c][][t]{0.8cm}{\centering \textbf{WT}} &
  \parbox[c][][t]{0.8cm}{\centering \textbf{TC}} &
  \parbox[c][][t]{0.7cm}{\centering \textbf{ET}} &
  \parbox[c][][t]{0.7cm}{\centering \textbf{WT}} &
  \parbox[c][][t]{0.7cm}{\centering \textbf{TC}} \\
  \hline
  \parbox[c][][t]{1.6cm}{\raggedright \textbf{Mean}}& 
  \parbox[c][][t]{0.8cm}{\centering {0.81}} &
  \parbox[c][][t]{0.8cm}{\centering {0.89}} &
  \parbox[c][][t]{0.8cm}{\centering {0.84}} &
  \parbox[c][][t]{0.8cm}{\centering {0.84}} &
  \parbox[c][][t]{0.8cm}{\centering {0.88}} &
  \parbox[c][][t]{0.8cm}{\centering {0.83}} &
  \parbox[c][][t]{0.8cm}{\centering {0.99}} &
  \parbox[c][][t]{0.8cm}{\centering {0.99}} &
  \parbox[c][][t]{0.8cm}{\centering {0.99}} &
  \parbox[c][][t]{0.7cm}{\centering {15.3}} &
  \parbox[c][][t]{0.7cm}{\centering {6.3}} &
  \parbox[c][][t]{0.7cm}{\centering {15.2}} \\
  \hline
  \parbox[c][][t]{1.6cm}{\raggedright \textbf{Std.}}& 
  \parbox[c][][t]{0.8cm}{\centering {0.20}} &
  \parbox[c][][t]{0.8cm}{\centering {0.11}} &
  \parbox[c][][t]{0.8cm}{\centering {0.24}} &
  \parbox[c][][t]{0.8cm}{\centering {0.21}} &
  \parbox[c][][t]{0.8cm}{\centering {0.13}} &
  \parbox[c][][t]{0.8cm}{\centering {0.25}} &
  \parbox[c][][t]{0.8cm}{\centering {.0004}} &
  \parbox[c][][t]{0.8cm}{\centering {.0007}} &
  \parbox[c][][t]{0.8cm}{\centering {.0007}} &
  \parbox[c][][t]{0.7cm}{\centering {69.5}} &
  \parbox[c][][t]{0.7cm}{\centering {28.9}} &
  \parbox[c][][t]{0.7cm}{\centering {63.9}} \\
  \hline
  \parbox[c][][t]{1.6cm}{\raggedright \textbf{Median}}& 
  \parbox[c][][t]{0.8cm}{\centering {0.85}} &
  \parbox[c][][t]{0.8cm}{\centering {0.92}} &
  \parbox[c][][t]{0.8cm}{\centering {0.92}} &
  \parbox[c][][t]{0.8cm}{\centering {0.92}} &
  \parbox[c][][t]{0.8cm}{\centering {0.92}} &
  \parbox[c][][t]{0.8cm}{\centering {0.94}} &
  \parbox[c][][t]{0.8cm}{\centering {0.99}} &
  \parbox[c][][t]{0.8cm}{\centering {0.99}} &
  \parbox[c][][t]{0.8cm}{\centering {0.99}} &
  \parbox[c][][t]{0.7cm}{\centering {1.4}} &
  \parbox[c][][t]{0.7cm}{\centering {2.8}} &
  \parbox[c][][t]{0.7cm}{\centering {2.2}} \\
  \hline
  \parbox[c][][t]{1.6cm}{\raggedright \textbf{25 quantile}}& 
  \parbox[c][][t]{0.8cm}{\centering {0.78}} &
  \parbox[c][][t]{0.8cm}{\centering {0.87}} &
  \parbox[c][][t]{0.8cm}{\centering {0.87}} &
  \parbox[c][][t]{0.8cm}{\centering {0.83}} &
  \parbox[c][][t]{0.8cm}{\centering {0.86}} &
  \parbox[c][][t]{0.8cm}{\centering {0.82}} &
  \parbox[c][][t]{0.8cm}{\centering {0.99}} &
  \parbox[c][][t]{0.8cm}{\centering {0.99}} &
  \parbox[c][][t]{0.8cm}{\centering {0.99}} &
  \parbox[c][][t]{0.7cm}{\centering {1.0}} &
  \parbox[c][][t]{0.7cm}{\centering {1.7}} &
  \parbox[c][][t]{0.7cm}{\centering {1.4}} \\
  \hline
  \parbox[c][][t]{1.6cm}{\raggedright \textbf{75 quantile}}& 
  \parbox[c][][t]{0.8cm}{\centering {0.93}} &
  \parbox[c][][t]{0.8cm}{\centering {0.95}} &
  \parbox[c][][t]{0.8cm}{\centering {0.96}} &
  \parbox[c][][t]{0.8cm}{\centering {0.96}} &
  \parbox[c][][t]{0.8cm}{\centering {0.96}} &
  \parbox[c][][t]{0.8cm}{\centering {0.97}} &
  \parbox[c][][t]{0.8cm}{\centering {0.99}} &
  \parbox[c][][t]{0.8cm}{\centering {0.99}} &
  \parbox[c][][t]{0.8cm}{\centering {0.99}} &
  \parbox[c][][t]{0.7cm}{\centering {2.2}} &
  \parbox[c][][t]{0.7cm}{\centering {4.8}} &
  \parbox[c][][t]{0.7cm}{\centering {3.8}} \\
  \hline
  \multicolumn{13}{c}{\parbox[c][][t]{10.9cm}{\centering ET - enhancing tumour, WT - whole tumour, TC - tumour core.}}\\
  \end{tabular}}
% \vspace{-1.5em}
\label{table:test_res}
\end{table*}

\section{Discussion and conclusions}
\label{sec:disc_conc}

In this work, we proposed an end-to-end automated tumour segmentation method using a triplanar ensemble architecture of 2D U-Nets. Our method segmented ET, WT and TC sub-regions of the tumour with dice values of 0.83, 0.93 and 0.87 on the training dataset. On an independent unlabelled validation dataset, our method achieved Dice values of 0.77, 0.89 and 0.77 for ET, WT and TC sub-regions respectively. On the BraTS'20 unseen test dataset, our method achieved the Dice values of 0.81, 0.89 and 0.84 for the ET, WT and TC regions respectively.

\medskip
Studying the effect of individual components on segmentation performance aided in better understanding the proposed method. For all the tissue classes, a single axial network performs the worst, probably due to the lack of contextual information from the contiguous slices. The triplanar architecture provides better performance than the individual 2D networks with higher Dice and sensitivity values. Moreover, we achieved significant improvement in the performance metrics with the addition of a specific TC network, especially for the TC class. The WT segmentation improves with the addition of each component and the significantly lower H95 values indicate a more precise tumour segmentation.

\medskip
From the results on the training and validation data, performance for the WT is higher than for the other two classes, indicating that the method segments the cumulative tumour region (including the edematous/invaded region) with higher accuracy than the differentiation between core and edema. However, a few misclassified tumour core regions were due to cases where the ET class is very small (as indicated by white arrows in fig.~\ref{fig:tr_res_vis}).  This results in ET either being incorrectly predicted (by the network) or falsely relabeled (in the post-processing step) as a part of the NCR/NET class. This substantially affects the Dice score for the ET class for the subject. In general, we observed that applying generalisable/consistent prior information or post-processing operations for TC and ET classes was not possible due to the wide range of tumour characteristics and variation in ground truth, which presented a major challenge for the segmentation task. Interestingly, for the TC region, our method performed better on the training and the unseen test datasets when compared to the validation dataset, as observed from the higher values of the metrics in tables~\ref{table:tr_res} and \ref{table:test_res} as compared to those in table~\ref{table:val_res}. The higher values on the training dataset could be due to the fact that the cross-validation results are generally more prone to over-fitting, and hence are less reliable when compared to the results on the unseen validation data. However, our method also provided consistently good performance on the unseen test dataset (table~\ref{table:test_res}).
Since this was an unseen dataset, we cannot exclude the possibility that the tumour characteristics and image intensity profiles of the test dataset could be more similar to those of the training dataset.
%, indicating the possibility that the tumour characteristics and image intensity profiles of the test dataset could be similar to those of the training dataset. % than those of the validation dataset. %robustness of the model to variation in the image characteristics and acquisition protocols. Since this was an unseen dataset, we cannot exclude the possibility that the tumour ...training dataset.

\medskip
On the BraTS'20 unseen test dataset, our method achieved  higher Dice values compared to the validation Dice values, obtaining an evaluation score that was equal 5$^{th}$ highest and 10$^{th}$ place in the overall ranking in the challenge. As an indirect comparison with existing methods using different validation datasets, our method achieved a Dice value of 0.89 (on the BraTS'20 validation dataset) for the WT class, comparable to the top-ranking methods of BraTS'17-19 ($\sim$0.90) \cite{kamnitsas2017efficient}, \cite{wang2017automatic}, \cite{myronenko20183d}, \cite{isensee2018no}, \cite{jiang2019two} and a Dice value of 0.77 for the ET class, on par with top-ranking methods of BraTS'17 ($\sim$0.78) \cite{kamnitsas2017efficient}, \cite{wang2017automatic}. It is worth noting that even though the previous BraTS challenges used data from different subjects, this indirect comparison could be quite useful in determining the potential of our method, since the comparison involves the same task and type of data.

\medskip
Summarising, our proposed triplanar ensemble method achieves accurate segmentation of whole tumours and their sub-regions on brain MR images from multimodal data of the BraTS'20 challenge. After the challenge, we will make our method publicly available as a Docker container that could be used as an independent tumour segmentation tool. Future directions include further improvement of tumour sub-region segmentation by leveraging the salient features (e.g. using attention networks). 

\section*{Acknowledgements}
\label{sec:acks}

The authors of this paper declare that their method for the BraTS'20 challenge has not used any pre-trained models nor additional datasets other than those provided by the organizers.  This work was supported by Wellcome Centre for Integrative Neuroimaging, which has core funding from the Wellcome Trust (203139/Z/16/Z).  VS is supported by Wellcome Centre for Integrative Neuroimaging (203139/Z/16/Z). LG is supported by the Oxford Parkinson’s Disease Centre (Parkinson’s UK Monument Discovery Award, J-1403), the MRC Dementias Platform UK (MR/L023784/2), and
the National Institute for Health Research (NIHR) Oxford Health Biomedical Research Centre (BRC). MJ is supported by the National Institute for Health Research (NIHR), Oxford Biomedical Research Centre (BRC) and Wellcome Trust
(215573/Z/19/Z). The computational aspects of this research were supported by the Wellcome Trust Core Award (203141/Z/16/Z) and the NIHR Oxford BRC. The views expressed are those of the authors and not necessarily those of the NHS, the NIHR or the Department of Health.

\end{document}